# Analysis of Urban Visitor's Walkability Based On Mobile Data: The Case of Daejeon, Korea

Mi Chang[a] , Gi-bbeum Lee[a] , and Ji-Hyun Lee[a]*

[a]*Graduate School of Culture Technology, KAIST, Daejeon, The Republic of Korea;*

E-mail address: jihyunlee@kaist.ac.kr (Ji-Hyun Lee)

**Mi Chang** is a Ph.D. student in Information-based Design Lab at KAIST. Also, she is an engineering researcher and product designer. She received M. Design, and B. Eng. in Electronic Communication Engineering. And she worked in LG electronics to develop TV.

**Gi-bbeum Lee** is a Ph.D student in Information-based Design Lab at KAIST. Her research interests center around human-AI interaction, digital placemaking, and remote collaboration. She is currently investigating the use of AI in future education and community design.

**Ji-Hyun Lee** is a Professor at the Graduate School of Culture Technology in KAIST. She received her Ph.D. in School of Architecture (Computational Design) at Carnegie Mellon University writing a thesis about integrating housing design and case-based reasoning.

**Declaration of competing interest**

The authors declare that they have no known competing financial interests or personal relationships that could have appeared to influence the work reported in this paper

**Acknowledgments**

This work was supported by National Research Foundation of Korea Grant (NRF) funded by the Korean Government [NRF-PM Support Program].

Highlights

- Data-driven walkability measurement method proposed
- Network analysis on mobile-based visitor data structurally evaluates walkability
- Point-of-interest (POI)-cell network analyzed through edge weight
- Understand city context through POI analysis of nine extracted communities
- Determine potential POI demand by visitor's walkability-based division

# Analysis of Urban Visitor's Walkability Based On Mobile Data: The Case of Daejeon, Korea

**Abstract**

The walkability of destinations plays a decisive role in obtaining a sense of place for urban visitors. For an improved walkability-based visit experience, the city structure is analyzed by dividing it into small units, starting with the destination. However, existing studies focus on the walkability of residents by administrative district and use limited contextual information and structural analysis tools. Therefore, this study aims to structurally understand walkability through network analysis with mobile-based visitor data collected from 822 point-of-interest (POI) in Daejeon. In this paper, the actual walking distance between 95,817 cells with visitors and POI is measured to calculate the edge weight and the influence of cells on POI. We analyze the POI-cell network through edge weight, extract nine communities, understand the city context through POI analysis and derive the following findings. Firstly, it is possible to determine the potential demand by the visitor's walkability-based division. Second, our walkability measurement method follows a bottom-up approach, starting with a small unit cell, understanding the entire city as a community extraction model, and possibly extending to other cities in Korea. Third, POI-centered structural analysis is possible using the number of visitors and distance in the proposed network analysis method.

## 1. Introduction

Research on visitors and mobility to city destinations is an essential topic in tourism study (Joppe, 2003; Pappalepore et al., 2010; Hoffman, 2008) and crucial for urban tourism planning (Russo & Richards, 2016; Wise, 2015). Visitors often walk or use some modes of transportation to reach their destination. Walking, especially, is a significant factor for residents and visitors. Because people seek an intense and varied experience in a short period in a compressed space, an excellent place to walk is fundamental for a visit (Clavé, 2018). As the visitor moves toward a destination on foot, it creates a sense of place (Xiao & Wei, 2021; Hall & Ram, 2018). The walkability determines the density of visitors' experience in the city (Marzbani, 2020; Griffin & Hayllar, 2009; Budruk et al., 2008; Kothencz & Blaschke, 2017). Therefore, it is important to analyze the walkability of the destination centering on the visitor.

The visitor context and impact on the destination can be determined by analyzing the city's spatial structure (Sugimoto et al., 2019). The '*15-minute city*' concept intends to understand the spatial structure of a city by dividing it into small units (Gaglione, 2021; Abdelfattah et al., 2022). The concept comes from the historical background of proximity and walkability mentioned by Row & Jacobs (1962) when pursuing a city where daily life is within walking distance, and the movement of residents is minimized within the neighborhood. Walkability is essential in understanding a city's physical structure and connecting people's social and personal quality of life (Leslie et al., 2007; Forsyth, 2015).

Though several studies have addressed methods to improve walkability for urban planning and tourism enhancement, three limitations exist. First, walkability studies from a visitor perspective are lacking. Ashworth and Page (2011) are even surprised by the lack of research despite the importance of walkability for visitors. Though visitors form a new destination image (Yadi & Putra, 2021) and sense of place (Marzbani, 2020) when walking to the destination, most studies are resident-biased. Second, there are practical limitations in the destination-oriented walkability analysis method. The current census in Korea conducts administrative district-based quinquennial surveys of visiting population (Lee et al., 2020). This method requires visiting individual houses, which is expensive and time-consuming (Lee et al., 2018). In addition, although the visitor considers the destination rather than the administrative district, the studies are conducted at the administrative district level. Moreover, although there are 40 officially designated attractions in Daejeon, the tools developed from a destination viewpoint to verify them are lacking. Third, even if there is a data-based walkability measurement method, it cannot be applied directly to Korea without understanding the context. Though various generalization methods, such as walk score, exist (Brown et al., 2013), the opportunity to see characteristic differences and details of Korean cities could be overlooked if applied collectively. For example, in Korea, the distance between administrative districts is close. Even when several places have the same administrative district, there are cases where interactions with places outside the administrative district exceed. Moreover, in places with many alleys and narrow roads, the difference between the straight and actual walking distance can be significant. It is necessary to divide the district with a walkability-based new concept outside the administrative district to reflect this local context.

Therefore, this study aims to structurally understand destination-oriented walkability through network analysis using mobile-based visitor data in Daejeon, a specific city in Korea, and analyze the city context through the influence and actual walking distance between destination and cell spatial unit. We collect point-of-interest (POI) frequently visited and cell-based visitor data, and measure the actual walking distance between POI and cell. The measured distance and the number of visitors on the cell are reflected in the edge weight for network visualization. Through this network, we extract the communities and analyze the community-related POI based on the walkability of visitors. The contribution of this study is three-fold. First, the potential demand, the key to urban tourism, can be identified by understanding the units of the visitor's walkability-based area in Daejeon. Second, the proposed walkability measurement method is a bottom-up method based on actual data, starting with a small spatial cell, structurally understanding the entire city, and possibly expanding to other cities in Korea. Third, the destination-oriented structural analysis is performed by connecting the actual walking distance and the number of visitors to the network.

The remaining paper is organized as follows: A review of similar works is presented in *Literature Review*; *Data and Methods* describe the analysis framework and details of the methods; *Results* on network analysis, community detection, and POI analysis are presented next, followed by a *Discussion* on the relevance of the proposed methods. Finally, the *Conclusion* concludes the paper by stating the study's implications, limitations, and future work.

## 2. Literature review

### 2.1. Walkability and Visitors

Walking is a social activity and a process by which pedestrians relate to various environmental elements of a city (Vale et al., 2016). Designing urban environments densely to enhance walkability allows residents or visitors to effectively access essential facilities and services without relying on transportation such as cars (Thornton et al., 2022). In addition, walkability is an essential factor contributing to a sustainable city, positively influencing residents' happiness index (Kim et al., 2019), social cohesion, civic engagement, public health, and traffic congestion (Hanibuchi et al., 2012; Vale et al., 2016). Because walkability is deeply related to the activities occurring in a city, it affects residents and visitors (Clavé, 2018). Walkability is significant for visitors to make an active tour and enhance their experience in the city (Freeman et al., 2013; Vale et al., 2016). Marquet (2020) states that among the built environment characteristics in the travel context, the level of walkability is related to the setting of a travel destination. Ujang (2014) also mentions that for tourist attractions to be reborn as tourist-friendly places in cities, place attachments should be increased by improving walkability.

Various methods exist to measure walkability (Rafiemanzelat et al., 2016), including the Walk Score index as the quantitative indicator (Ram & Hall, 2018; Zhang et al., 2022; Koohsari et al., 2018; Hirsch et al., 2013; Brown et al., 2013). Kim et al. (2019) also proposed a Walk Score algorithm-based surveying method for measuring walkability in dense cities such as Seoul. Hanibuchi et al. (2012) calculated the Walk Score by measuring the built environment based on the population density to quantify the walkability in an Asian environment. Such accessibility measurement methods are divided into place-based and individual-based methods. An example of the place-based method is the Pedestrian Environment Index developed by Peiravian (2014), which quantifies the urban neighborhood friendliness of pedestrians using population, land-use, commercial density, and intersection density. Using Yokohama residents as subjects, Hino et al. (2022) proposed an individual-based walkability index that quantifies the number of steps taken and the accessibility to various urban amenities.

However, most walkability measurement studies focus only on factors such as distance to a specific facility and environment, indicating a bias toward place-based methods. Although walkability is important to visitors, systematic analysis is limited to understanding their level of existence around the destination and the actual walking distance to the destination. Though individual-based accessibility measurement evaluates accessibility by considering various variables influencing visitor decisions (Vale et al., 2016), there is a lack of walkability studies that focus on the distance, destination, and visitors.

*2.2. Understanding Urban Structure and Network analysis*

Recognizing the continuously changing spatial structure is needed to understand the space in a city (Helsley & Strange, 2007; Dadashpoor & Malekzadeh, 2020). The spatial distribution of cities is nonuniform with distinct characteristics and has continued development filled with uncertainty (Vertesi et al., 2019). The neighborhood is an appropriate concept to understand the structure of a city through walking, transportation, and infrastructure (Huang et al., 2019; Braver et al., 2022; Hendricks et al., 2018). According to Kissfazekas (2022), the neighborhood concept has been in existence for nearly 100 years and shares some similarities with the 15-minute city. Freeman et al. (2013) defined a neighborhood with a zip code and analyzed walkability. Hanibuchi et al. (2012) measured walkability by population density and defined the neighborhood as a radial 500 m with the built environment and census. Furthermore, virtual environment videos measure walkability-based neighborhood formation from a user perspective (Liao et al., 2022). Community detection using graph theory can effectively explain the urban structure based on neighborhood connectivity. For example, Shao et al. (2017) used community detection through social media data to define tourism districts in Huangshan city. Zhang et al. (2021) characterize the urban structure as community changes over time, targeting London's smart card data. In the same year, Hu et al. (2021) used geospatial data to understand urban space by community theme. Li & Zhang (2016) analyzed the spatial structure based on travel activities using the transportation system in a community.

However, there are limitations to applying the neighborhood analysis results in a top-down way to Korea. Although there is an attempt in a bottom-up way to analyze upper neighborhoods from non-work destination data (Ponce-Lopez & Ferreira, 2021), it is limited to analyzing the walking distance or the connectivity between clusters. Even if community detection is performed based on actual data, understanding the city's structure in the context of an urban visitor is challenging. Further research is needed to understand the bottom-up method's urban structure that fits Korea's context.

## 3. Data and Methods

The proposed analysis framework is shown in Fig. 1. First, we collect POI frequently visited in Daejeon and mobile-based visitor data. Because the location data format in the visitor data is in Korean standard UTM-K, it is converted into WGS84 for optimal analysis using the Python library. Second, we calculate the distance between POI and cells with visitors connected within 1 km. Third, the number of visitors in each cell is calculated to determine the accumulated visitors for November 2021. The visitor influence is determined as a weight according to the distance between POI and cell. Fourth, the network edge weight is determined using the calculated visitor and distance data, and the network is visualized. Fifth, we detect communities through modularity and compare their characteristics over time. Finally, the POI characteristics are determined based on the analyzed community characteristics.

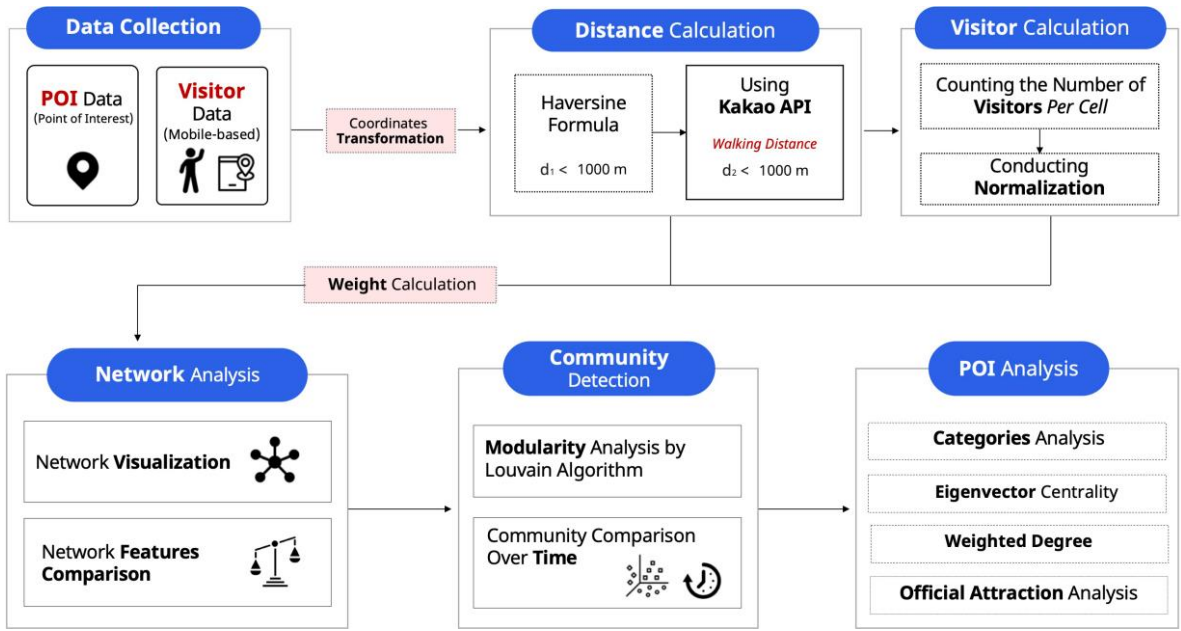

**Fig. 1** Proposed framework to analyze the relationship between POI and visitors.

*3.1. Study Area*

We chose *Daejeon* city, one of the most visited cities in Korea, as the study area. It is the fifth most significant metropolis in South Korea in terms of population (1.4 million) and size (539.53 km$^2$).[1] Located in the central region of the Korean Peninsula, Daejeon, meaning a large field, is ringed by mountain ranges and Daecheong Lake. Historians estimate that the name came into being 500 years ago[2] and has several government buildings and cultural places. It is called Korea's home of science and technology industries as it houses the Korea Advanced Institute of Science and Technology, Electronics and Telecommunications Research Institute, and Korea Aerospace Research Institute. The abundance of social, educational, and cultural facilities and natural surroundings attracts visitors annually. Besides, Daejeon serves as a transportation hub, connecting the capital and provinces. The expressways Honam and Gyeongbu Lines, having enormous traffic volumes, pass through the city. Daejeon's train station and distribution center are of significant scale,[3] and thousands of professionals and travelers visit the city in all seasons. The administrative divisions are divided into five regions: Dong-Gu, Jung-Gu, Seo-Gu, Yuseong-Gu, and Daedeok-Gu (Fig. 2). Yuseong-Gu has the largest area and the most prominent research institutes. Seo-Gu is the most populated because of the densely clustered government offices, commercial and residential buildings, and educational facilities. Daecheong Lake and transportation landscapes belong to Dong-Gu. Because each district has markedly different characteristics, visitors travel through various divisions depending on their purpose.

---

[1] https://kosis.kr/statHtml/statHtml.do?orgId=127&tblId=DT_127006_B006&conn_path=I2
[2] https://www.daejeon.go.kr/
[3] https://www.dsi.re.kr/board.es?mid=a10112000000&bid=0026

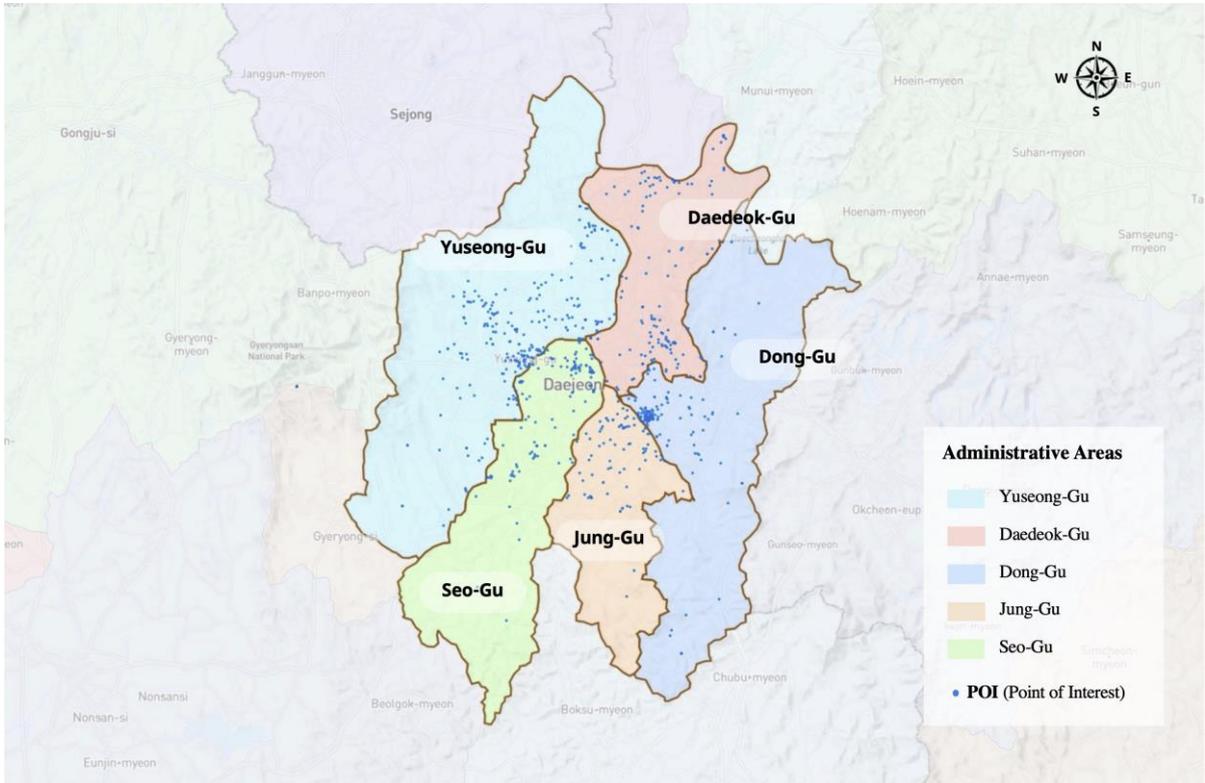

**Fig. 2** Daejeon city map; Five divisions of Daejeon city and POI frequented by visitors

*3.2. Datasets*

3.2.1. Mobile-based Visitor Data

We used the mobile-based visitor dataset for November 2021 collected by SK Telecom (SKT), one of the largest telecommunication companies in Korea. As of March 2022, they have approximately 30 million subscribers, accounting for 40% of the population.[1] The data is from a 50×50 m cell based on the mobile call location. The number of calls is snapshotted hourly. The number of 'visitors' is the number of people entering the cell outside the individual's residence and workplace based on the corresponding process. In this process, SKT estimates the 'Dong' where the subscriber stayed the most between 0:00-6:00 and 9:00-16:00 in the previous month (October) as the actual residence and workplace, respectively. When an individual visits a specific cell outside their residence and workplace, the number of visitors reflects the totalization weight to find the visiting population in the cell. Visitor data consists of a unique cell id, snapshot time, number of visitors, latitude, and longitude (Table 1). Data is included in the snapshot only when a visitor population is in the cell (Fig. 3). The measurement data of all small cells located in Daejeon is included in the spatial dimension. We collected 59,743,989 records in total from the 95,817 cells showing visitors.

---

[1] https://kosis.kr/statHtml/statHtml.do?orgId=127&tblId=DT_127006_B006&conn_path=I2

**Table 1.** Example of a mobile-based visitor data record

| Column | Data | Description |
| --- | --- | --- |
| cell_id | c4875 | Unique id of the cell |
| time | 20211101-12 | Year, month, date, time of the data |
| visitors | 10 | Number of visitors in the cell |
| Lat | 36.279981 | Latitude coordinate |
| Lon | 127.411476 | Longitude coordinate |

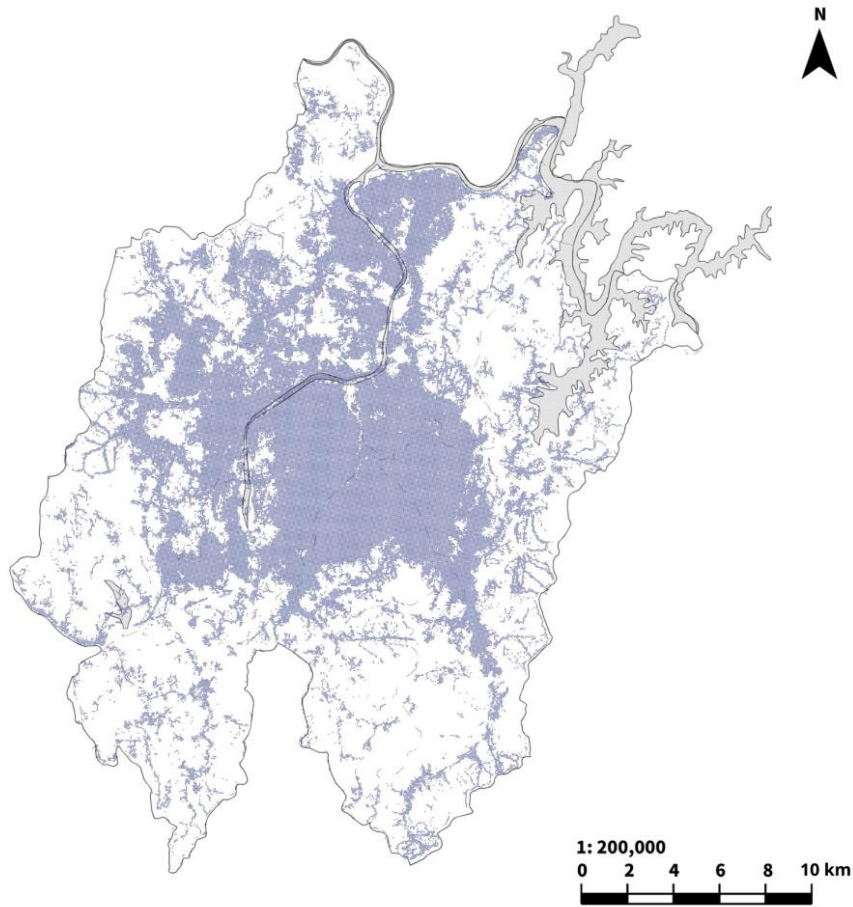

**Fig. 3** Cell distribution with visitors in Daejeon for November 2021

3.2.2. POI data

By collecting POI, we can understand the city's context based on the destination of visitors. The collection route is obtained from Google map, Naver map, Instagram, Facebook, and six Daejeon official homepages (one city hall [1], five district offices [2]), which Koreans most frequently use to find a POI. We collected 850 POI, including 40 tourist attractions officially designated by local governments. POI attributes include *poi_id*, *title*, *address*, *lat*, *lon*, *categories,* and *official_attraction*, as shown in Table 2.

The mobile-based visitor data is then used to verify whether the collected POI can be considered an actual POI. The visitors' data divided by period was accumulated based on cell id. As shown in Table 3, the average number of accumulated visitors per POI is approximately 1,941,884, and nearly 60% of POI have over 500,000 visitors. In addition, the POI visited by the most significant number of visitors is p89, a restaurant in Dunsan-Dong, Seo-Gu. However, there are 28 places with 0 visitors and are deleted from POI data. Finally, 822 POIs are determined, as identified in Fig. 2 and Table 4.

**Table 2.** Example of a POI record (the content in italic is translated from Korean)

| Column | Value | Description |
| --- | --- | --- |
| poi_id | p436 | Unique id of the POI |
| title | *Traditional Food Experience Center* | Name of the POI |
| address | *San 4-1, Musu-Dong Jung-Gu, Daejeon* | Address of the POI |
| Lat | 36.279981 | Latitude coordinate |
| Lon | 127.411476 | Longitude coordinate (Centroid) |
| categories | experience_center | Category name of POI (52 categories) |
| official_attraction | False | Whether included in the 40 tourist attractions designated by the local government |

**Table 3.** Number of visitors by POI

|  | Mean | Min | 25% | 50% | 75% | Max |
| --- | --- | --- | --- | --- | --- | --- |
| Number of Visitors | 1,941,884 | 0 | 73,356 | 951,541 | 3,024,857 | 9,777,765 |

---

[1] https://www.daejeon.go.kr/

[2] https://www.seogu.go.kr/
https://www.donggu.go.kr/
https://www.yuseong.go.kr/
https://www.djjunggu.go.kr/
https://www.daedeok.go.kr/

**Table 4.** Number of POI by Gu

| Name of Gu | Yuseong-Gu | Daedeok-Gu | Dong-Gu | Jung-Gu | Seo-Gu | Sum |
|---|---|---|---|---|---|---|
| Number of POI | 272 | 161 | 75 | 181 | 132 | 822 |

POI is divided into 52 categories and is listed in Table 5. Park is the most common, followed by café and cultural heritage. Characteristically, museums and bakeries, scarce in other cities, occupy a significant portion of the POI.

**Table 5.** POI all categories

| POI Categories | Park | Café | Cultural Heritage | Restaurant | Museum | Forest | Experience Center | Street | Bakery | Others |
|---|---|---|---|---|---|---|---|---|---|---|
| Count | 159 | 113 | 104 | 92 | 32 | 29 | 25 | 24 | 16 | 228 |
| Percentage | 19% | 14% | 13% | 11% | 4% | 4% | 3% | 3% | 2% | 28% |

The local characteristics of the extracted POIs are identified through POI-POI and POI-cell relationships. The degree of connection between two POIs is determined by the number of commonly connected cells and the distance between the POI and the shared cells. Two POI can share a cell when they exist within a radius of 1 km from the cell. Fig. 4 shows the degree of interconnection with the number of cells shared by all pairs of POIs. When the two POIs are located at almost the same location, up to 800 cells are shared.

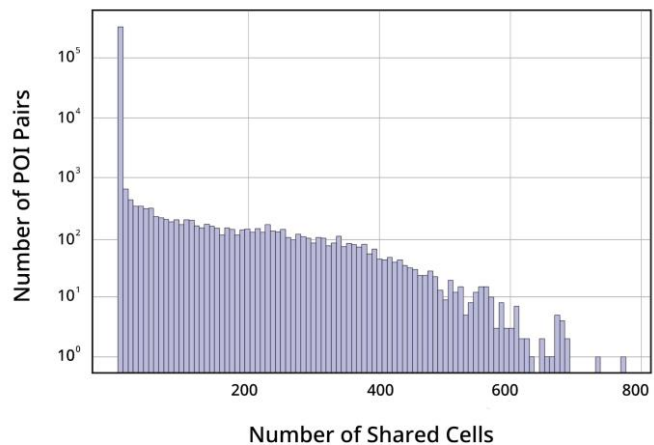

**Fig. 4** Density of POI-POI connections in Daejeon city

The POI-cell relationship is determined using the distance between their centroids. The closer the centroid of POI and the cell is, the stronger the connection between them. Moreover, the closer the POI and cell, the smaller the distance, even overlapping in extreme cases. Fig. 5 shows the degree to which 95,817 cells of mobile-based visitor data and 822 POIs have connections within 1 km walkability. POI density is closer to 1 when a cell is connected to many POIs and is close to zero when the number of connections is few. Cells with zero density are excluded from the data analysis (Fig. 5).

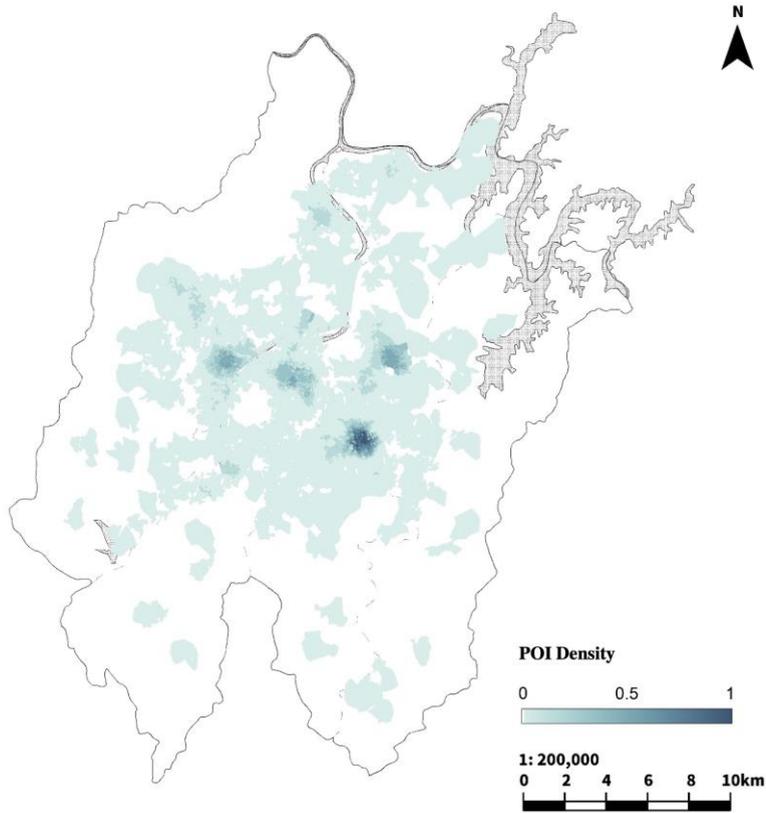

**Fig. 5** Density of POI-cell connections in Daejeon city

*3.3. Analysis Method*

3.3.1. Distance Calculation

The actual walking distance between POI and cell is then calculated. Considering a 15-minute city as a standard, walkability is 1 km based on a 4 km/h walking speed. In addition, we assume that the distance the visitors are willing to walk to a specific destination is approximately 1 km (Chu & Chapleau, 2008; Martínez & Viegas, 2013). Therefore, we store the POI-cell list only if the actual walking distance is within 1 km. Because counting all cases of POI and cells is time-consuming, we compute the linear distance between them to extract the first list within 1 km. Here, the straight-line distance is obtained using Python's *haversine*, and the number of cases reduces from over 40 billion to 720,000. Second, we measure the actual walking distance using the *Kakao* API, which provides map and navigation functions in Korea. The measured distance is the distance traveled by walking on the main road between the two points, as shown in Fig. 6. Accordingly, the elements in the list with a 1 km distance between POI and cell are reduced from 720,000 to 303,764.

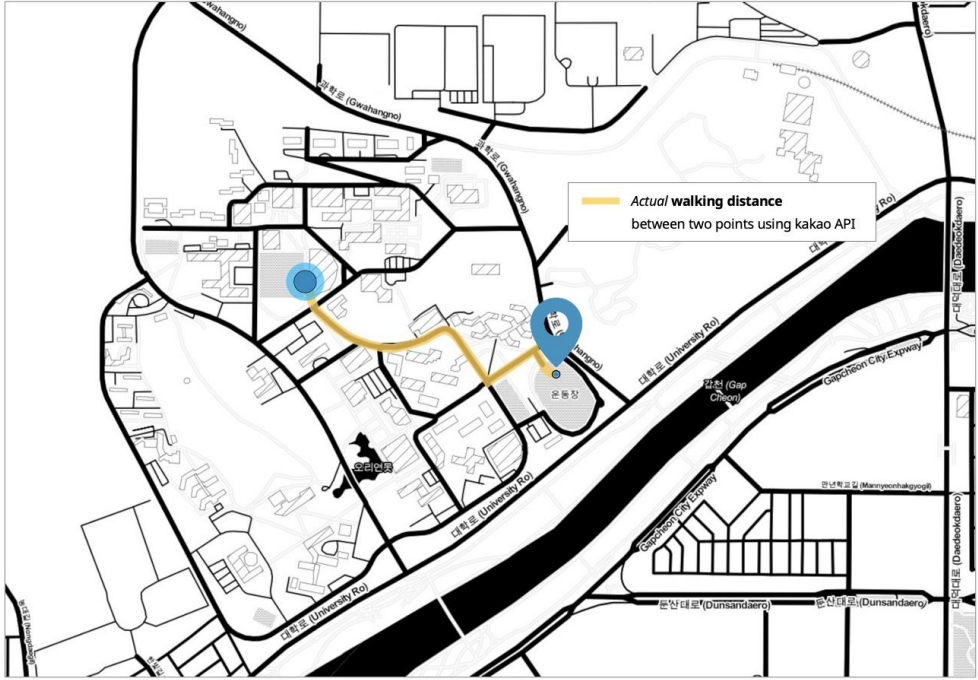

**Fig. 6** Example of calculating the walking distance between POI and cell.

3.3.2. Network Analysis

As shown in Fig. 7, there are two kinds of nodes: POI and visitors on cell. The network analysis is conducted using the distance between POI and cell calculated in subsection *Distance Calculation*. As in Equation (1), the edge weight of the network, $W_{edge}$ is defined by the distance between POI and cell and the number of visitors on cell. $N_v$ is the number of visitors in a cell and is an indicator of the influence of the cell on itself. Physical proximity $P_{pv}$ indicates the influence of the cell on the POI based on the distance between the POI and cell (Chang & Lee, 2021). $P_{pv}$ is normalized from zero to one; the closer the distance to zero, the closer the $P_{pv}$ to one. For example, suppose there are 10 visitors in a specific cell, i.e., $N_v$ = 10. If the distance between POI and cell is 200 m, $d_{max}$ is 1,000, $P_{pv}$ is 0.8, and $W_{edge}$ is 8 (10×0.8). Thus, the influence of cells on themselves through visitors ($N_v$) and toward POIs ($P_{pv}$) occupies an essential part in forming the network. All calculations were performed using *Pandas* and *NumPy* libraries in *Python* (3.7), while *Gephi* (0.9.2) was used for network visualization and statistical analysis.

$$W_{edge} = N_v \times P_{pv} = N_v \times \left(-\frac{d}{d_{max}} + 1\right) \tag{1}$$

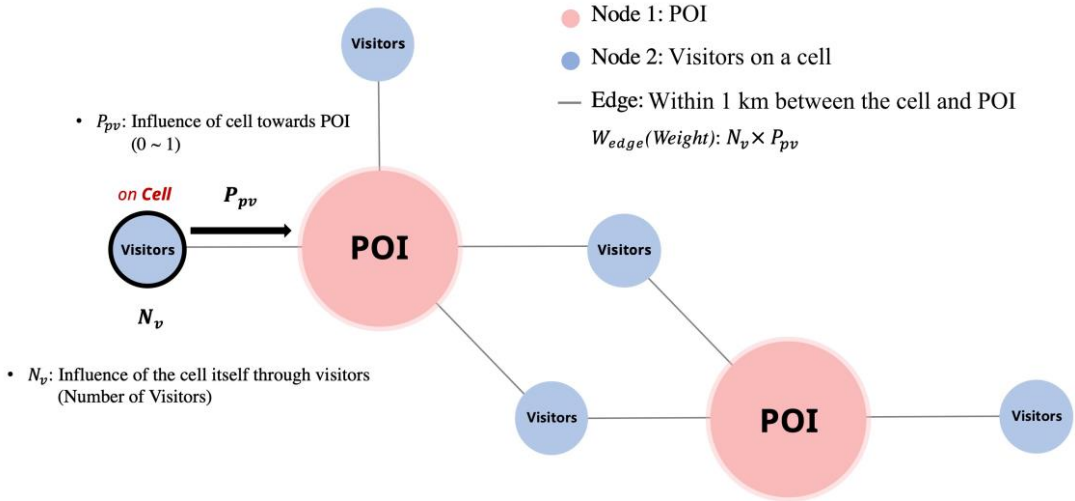

**Fig. 7** Description of network formation reflecting walkability and number of visitors.

3.3.3. Community Detection

In the POI-cell network, we analyze the community, such as the neighborhood, based on the walkability of visitors. The modularity value extracts the community through the Louvain algorithm, the most widely used community extraction method. The modularity of a network is expressed as a number between zero and one and indicates the network cluster compared to a randomly generated network. Communities are extracted for each modularity class, and those with more than 2,000 nodes are considered the leading community. Here, the resolution is set to 50 so that the primary communities occupy more than 70% of the network. The extracted central communities are compared and analyzed according to the number and category of POI, the number of visitors, and network characteristics.

3.3.4. POI Analysis

We analyze the relationship between POI and visitors from two perspectives: (1) within and (2) outside the community. First, in the internal analysis, POI with high values is analyzed using eigenvector centrality and weighted degree for each community. Here, the weighted degree means the sum of the edge's weight, and the eigenvector centrality measures how well the node is connected to an influential POI in the network. In addition, the current status of official attractions designated by Daejeon City and POI categories for each community are analyzed. Second, the current status of eliminated POI is analyzed by categories, using 'Gu' as an administrative district.

## 4. Results

### *4.1. Network Analysis and Community Detection*

As shown in Fig. 8, the POI-cell network is visualized by considering visitors in the region in November 2021. The network consists of 56,060 nodes and 245,274 edges. The average weighted degree is 9,299, and modularity is calculated as 0.78 (Table 6).

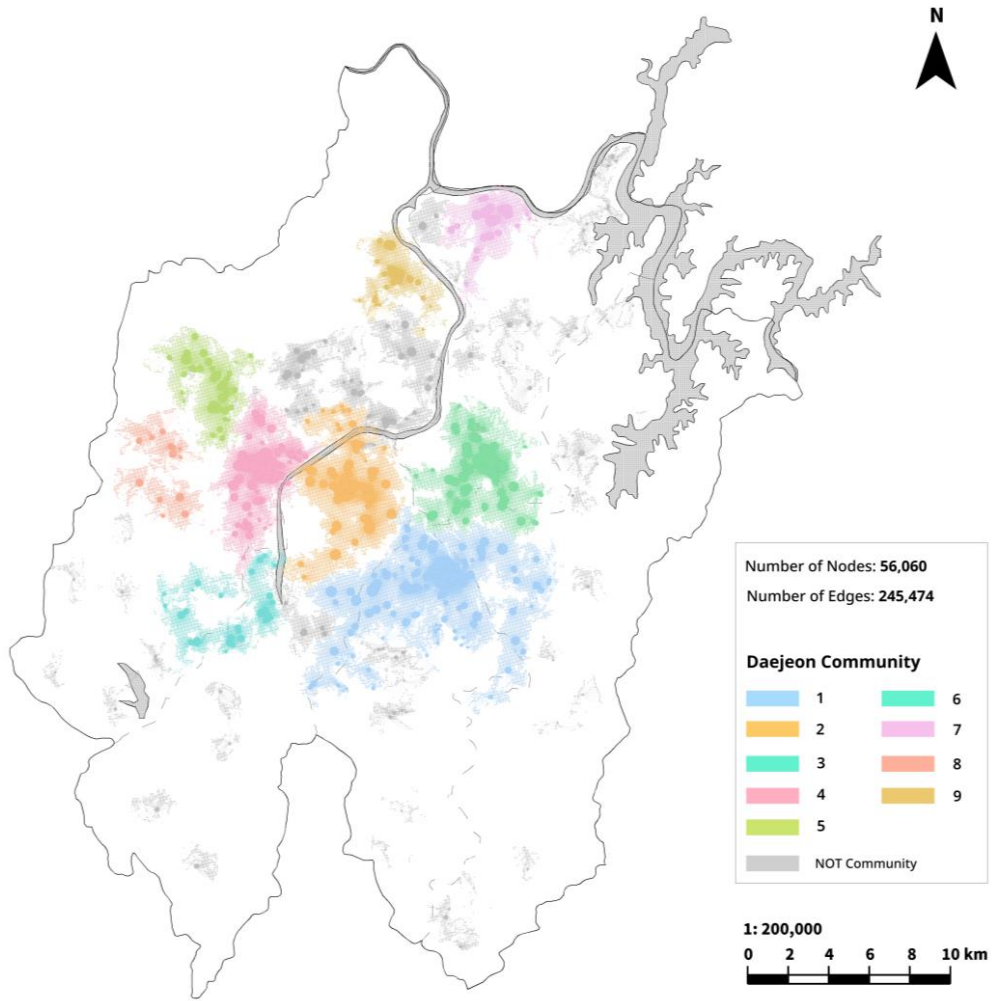

**Fig. 8** Communities detected in Daejeon city based on walkability and number of visitors.

**Table 6.** Statistical properties of POI and visitor's network

| Properties | Value |
|---|---|
| Node | 56,060 |
| Edge | 245,474 |
| Average Degree | 4.38 |
| Average Weighted Degree | 9,299 |
| Modularity | 0.78 |

Based on the modularity class, 80 communities are found. From them, communities with more than 2,000 nodes and at least a 3% ratio are extracted. The extracted nine communities are color mapped, as shown in Fig. 8. The node size indicates the degree of eigenvector centrality. In addition, the names of the communities are determined according to the community size.

The network characteristics and POI categories of the nine extracted communities were compared. A comparison of network characteristics, including the number of nodes and edges in each community, average weighted degree, and the number of POI, is presented in Table 7. The 246 POIs outside the nine communities will be analyzed in-depth in subsection *POI Analysis*.

**Table 7.** Network properties of communities in Daejeon

| Community Id | Number of Nodes | Number of Edges | Modularity | Average Weighted Degree | Percentage of Community | Number of POI |
|---|---|---|---|---|---|---|
| 1 | 12,691 | 64,606 | 0.61 | 12,851 | 22.64% | 171 |
| 2 | 6,696 | 35,900 | 0.64 | 20,813 | 11.94% | 84 |
| 3 | 5,871 | 38,140 | 0.65 | 12,164 | 10.470% | 85 |
| 4 | 4,760 | 34,301 | 0.55 | 17,494 | 8.49% | 87 |
| 5 | 3,273 | 13,692 | 0.62 | 4,001 | 5.84% | 42 |
| 6 | 3,163 | 10,911 | 0.71 | 3,680 | 5.64% | 39 |
| 7 | 2,497 | 8,870 | 0.57 | 4,498 | 4.45% | 26 |
| 8 | 2,198 | 4,157 | 0.71 | 910 | 3.92% | 14 |
| 9 | 2,173 | 8,475 | 0.43 | 4,908 | 3.88% | 28 |
| Sum | 43,322 | 219,052 | - | - | 77.27% | 576 |
| Not Community | 12,738 | 26,422 | - | - | 22.73% | 246 |

Community 1 is the largest community and has the most number of POI. Moreover, it has the second-highest average number of visitors (Table 8). Cafes, parks, and restaurants were the most distributed POI categories (Fig. 9).

Community 2 has the highest average weighted degree and the average number of visitors per POI. However, the number of nodes and edges only occupies about half of that of community 1. From a network viewpoint, the connectivity is somewhat inferior but has a high ratio of restaurants, hotels, and department stores.

Community 3 has more edges than community 2 but a smaller average weighted degree. Hence, the number of visitors is relatively low, and the influence of visitors on cells toward a POI is low. This community has the only high ratio of cultural heritage among POI categories.

Community 4 has weak clustering modularity compared to other communities. Most POIs are concentrated in Yuseong-Gu, and parks have the highest proportion among the POI categories. Compared to community 3, the average weighted degree is lower, but the average number of visitors in POI is relatively large. Similarly, in community 5, the number of edges is lower than that of community 4, but the modularity is slightly higher. Communities 6 and 8 have the highest modularity. However, these two communities have the smallest number of visitors per POI.

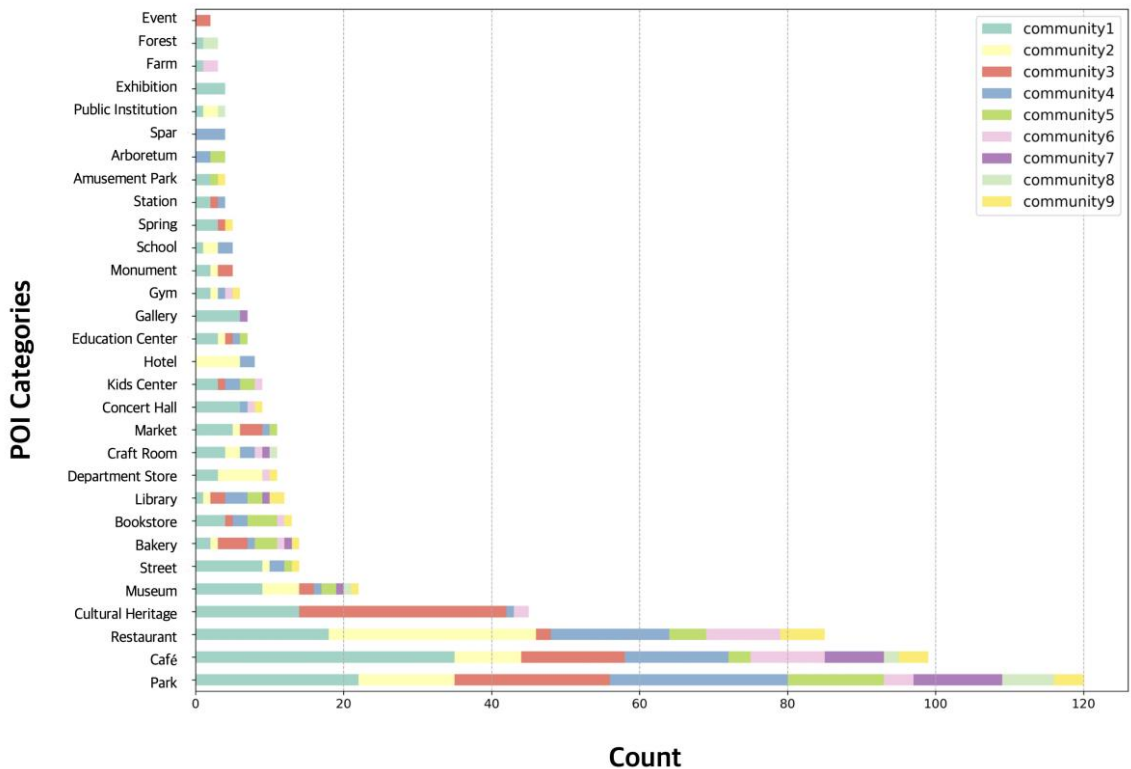

**Fig. 9** Category-wise distribution of POI by communities

**Table 8.** Characteristic comparison based on POI of communities

| Community Id | Average Number of Visitors Per POI | Number of POI | | | | |
|---|---|---|---|---|---|---|
| | | Yuseong-Gu | Daedeok-Gu | Dong-Gu | Jung-Gu | Seo-Gu |
| 1 | 3,123,676 | - | - | 26 | 142 | 3 |
| 2 | 5,222,800 | 10 | - | - | - | 74 |
| 3 | 2,473,647 | - | 76 | 9 | - | - |
| 4 | 2,814,205 | 85 | - | - | - | 2 |
| 5 | 1,063,479 | 42 | - | - | - | - |
| 6 | 841,621 | 11 | - | - | - | 28 |
| 7 | 1,174,154 | - | 26 | - | - | - |
| 8 | 396,285 | 14 | - | - | - | - |
| 9 | 1,098,108 | 28 | - | - | - | - |
| Sum | - | 190 | 102 | 35 | 142 | 107 |

*4.2. Comparison*

By comparing the relationship between POI and visitor by date and time, we understand the context of the city from a visitor perspective. We selected Wednesday, November 3, 2021, and Saturday, November 6, 2021, to compare the relationship between POI and visitors by time and day (Table 9). A significant difference between weekdays and weekends is that even though the number of cells is large on weekdays, the number of visitors is small. Here, a cell is counted only when there are visitors; if there is even one visitor on the cell, the number of cells goes up.

**Table 9.** Comparative analysis of cell and visitor according to date and time

| Date | Time | Number of Cells | Visitors on Cells | | |
|---|---|---|---|---|---|
| | | | Sum Total | Mean | Max |
| Weekday | All Day | 91,404 | 7,631,404 | 83.49 | 6,154 |
| (November 3) | 9 a.m. | 86,792 | 414,684 | 4.77 | 303 |
| | 1 p.m. | 87,212 | 462,012 | 5.30 | 403 |
| | 6 p.m. | 84,571 | 407,776 | 4.82 | 453 |
| | 11 p.m. | 82,453 | 194,110 | 2.35 | 123 |
| Weekend | All Day | 85,245 | 8,746,376 | 102.60 | 10,610 |
| (November 6) | 9 a.m. | 82,041 | 385,555 | 4.70 | 315 |
| | 1 p.m. | 82,256 | 526,686 | 6.40 | 837 |
| | 6 p.m. | 80,338 | 479,331 | 5.97 | 765 |
| | 11 p.m. | 78,350 | 282,833 | 3.61 | 406 |

The total number of cells and visitors increased during the weekend. The number of visitors is spread out on a weekday and is relatively small. On weekends, the visiting population is concentrated in a specific place, and the number of visitors is relatively large. On each date, 1 p.m. has the most significant number of cells and the largest number of visitors. However, the total number of visitors and maximum visitors per cell is 1.15 and 1.72 times higher, respectively, on weekends.

The difference in visitor behavior on weekdays and weekends in POI is further illustrated in Fig. 10. The node size represents the number of visitors connected to the POI, and the color represents the top 10 categories. Fig. 10(a) shows the visitor status of POI during the weekday. The number of visitors in the circled part A is large. Part A is Seo-Gu, where the famous department stores and the largest flower wholesale market in Daejeon are located. In addition, the 10 POIs with the most visitors are in this area. As shown in Fig.10 (b), the pattern of visitors changes on weekends. Overall, the number of visitors to POI increased, especially in parts B and C. These areas are Jung-Gu and Dong-Gu, having the largest train station and bus terminals. Marked regions A, B, and C show significant differences in POI between weekends and weekdays.

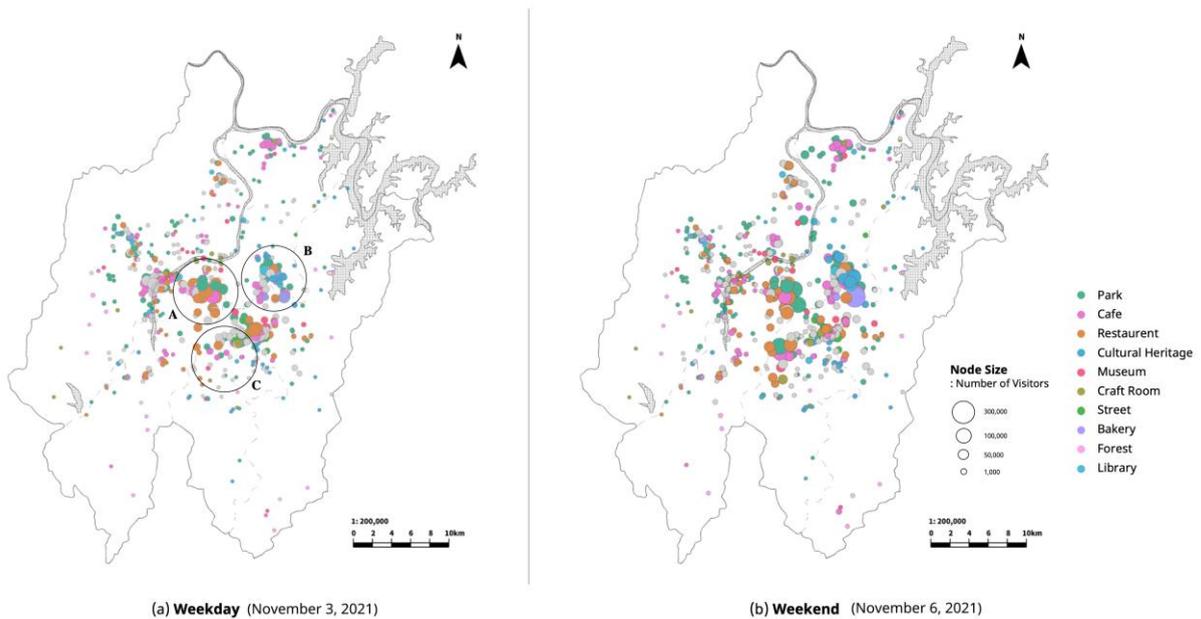

**Fig. 10** Changes in the number of visitors at POI on a weekday and weekend.

*4.3. POI Analysis*

The general destination-oriented behavior was investigated by comparing the patterns of POI-centered visitors on weekdays and weekends. Expanding further, we conduct a POI analysis including 40 official attractions designated by the Daejeon City administration to understand the destination-oriented context in detail. As previously described, the POI analysis is divided into two steps. The first involves analysis of the extracted communities and the second analyses outlier POIs in the network and nine communities.

The result of analyzing the POI based on the extracted community is presented in Table 10. For each community, the top 30 POIs are selected according to weighted degree and eigenvector centrality to extract the POI with the maximum value for the top five categories.

Community 1 is closely related to parts B and C (see subsection *Comparison*) and is frequented by visitors on weekends. This area is the original downtown of Daejeon and is located between two expressway stations. Hence, restaurants and cafes are active in the vicinity. Moreover, it is changing as public institutions are being relocated to other places along with developing new towns. Visitors and cultural artists are taking over the city as public institutions are being relocated. Moreover, as the customer base changes, stores are also changing. As shown in Table 10, the five categories of 30 POIs are cafés, restaurants, streets, bookstores, and bakeries in that order. The street category p431, a street for middle and high school students, and the landmark bakery p138 were extracted. In the eigenvector centrality analysis part, café and restaurant still occupy high categories, and p509 and p514 associated with gallery and exhibition, respectively, are shown to have a strong connection with POI. Among 40 official attractions, eight POIs (p14, p136, p138, p143, p148, p153, p374, and p464) belong to this community. However, only p138 (bakery) is extracted as the top POI; the remaining seven were low in degree and centrality.

Community 2 is closely related to the part A area (mentioned in subsection *Comparison*) and is more active during the weekdays. It has been established as a new downtown area due to apartment development and relocation of public institutions in the last 35 years. The department stores and hotels exist in the upper category (Table 10). The result indicates that many visitors to p692 (department store) have been with the new downtown since its development. Because there are many public institutions around it, restaurants, cafes, hotels, and parks with good walkability are available. For eigenvector centrality analysis, p761, a sizeable wholesale flower market, having the highest weighted degree and relatively many connections with other POIs, is extracted. In addition, this community includes four official attractions (p125, p128, p138, and p155). These POIs were outside the top 50 because of their low degree and centrality.

**Table 10.** POI Analysis: Extraction of top five categories of 30 POIs with the highest weighted degree and eigenvector centrality

| Community Id | POI analysis based on weighted degree | | | POI analysis based on eigenvector centrality | | |
|---|---|---|---|---|---|---|
| | Categories | POI | Weighted Degree | Category | POI | Eigenvector Centrality |
| 1 | Café | p401 | 2,985,326 | Café | p381 | 0.78255 |
| | Restaurent | p784 | 3,721,854 | Restaurent | p748 | 0.85503 |
| | Street | p431 | 2,684,708 | Gallery | p509 | 0.76913 |
| | Bookstore | p400 | 2,353,204 | Street | p431 | 0.76779 |
| | Bakery | p138 | 1,938,833 | Exhibition | p514 | 0.73289 |
| 2 | Restaurent | p68 | 3,697,963 | Restaurent | p87 | 0.81074 |
| | Café | p767 | 3,514,770 | Department Store | p762 | 0.75034 |
| | Department Store | p692 | 3,220,815 | Café | p768 | 0.74899 |
| | Park | p61 | 3,156,194 | Park | p105 | 0.73691 |
| | Hotel | p104 | 2,098,253 | Flower Shop | p761 | 0.72617 |
| 3 | Park | p560 | 1,843,887 | Cultural Heritage | p146 | 0.90201 |
| | Café | p584 | 1,829,927 | Park | p560 | 1.00000 |
| | Cultural Heritage | p629 | 1,105,771 | Café | p592 | 0.97315 |
| | Bakery | p585 | 1,871,951 | Bakery | p608 | 0.91007 |
| | Restaurant | p637 | 1,869,821 | Library | p596 | 0.81476 |
| 4 | Park | p231 | 3,195,099 | Park | p231 | 0.87517 |
| | Café | p300 | 2,675,779 | Restaurant | p705 | 0.96107 |
| | Spar | p142 | 3,063,242 | Café | p300 | 0.90201 |
| | Restaurant | p705 | 2,821,992 | Spar | p194 | 0.85369 |
| | Hotel | p709 | 3,049,990 | Hotel | p709 | 0.86040 |
| 5 | Park | p191 | 1,139,300 | Park | p213 | 0.75973 |
| | Bookstore | p330 | 1,105,854 | Restaurant | p825 | 0.55436 |
| | Restaurant | p825 | 770,518 | Bookstore | p330 | 0.71812 |
| | Bakery | p308 | 711,122 | Bakery | p308 | 0.66309 |
| | Arboretum | p205 | 164,947 | Street | p228 | 0.66980 |
| 6 | Restaurant | p719 | 728,391 | Café | p736 | 0.60403 |
| | Café | p736 | 865,629 | Restaurant | p719 | 0.55571 |
| | Park | p216 | 231,377 | Park | p216 | 0.44832 |
| | Concert Hall | p98 | 665,040 | Cultural Heritage | p348 | 0.54899 |
| | Cultural Heritage | p348 | 509,479 | Concert Hall | p98 | 0.44564 |
| 7 | Park | p550 | 948,862 | Park | p543 | 0.83758 |
| | Café | p580 | 1,814,398 | Café | p576 | 0.83222 |
| | Craft Room | p635 | 407,274 | Craft Room | p635 | 0.35705 |
| | Library | p625 | 76,839 | Museum | p578 | 0.50604 |
| | Museum | p578 | 64,787 | Library | p625 | 0.30738 |
| 8 | Park | p215 | 274,718 | Park | p237 | 0.58658 |

|   | Forest | p181 | 310,994 | Forest | p181 | 0.54362 |
|   | Café | p782 | 64,017 | Café | p782 | 0.35168 |
|   | Observatory | p169 | 107,273 | Craft Room | p226 | 0.48322 |
|   | Craft Room | p226 | 108,845 | Observatory | p169 | 0.42685 |
| 9 | Restaurant | p817 | 778,777 | Resturent | p67 | 0.69530 |
|   | Park | p246 | 750,676 | Park | p186 | 0.70336 |
|   | Café | p298 | 19,953 | Café | p298 | 0.30738 |
|   | Library | p340 | 608,631 | Library | p340 | 0.50201 |
|   | Department Store | p292 | 371,109 | Bakery | p293 | 0.67383 |

*All POI are considered for communities with less than 30 POI

In community 3, cultural heritage is widely distributed among all categories of POI. p146 is the only official attraction in this community and has a relatively high eigenvector centrality of 0.90201, but the weighted degree in the community is 26th. Although physically connected to influential POIs, the node's actual influence is small. In addition, this community is an area with a high linkage between parks. Among parks, p560 has an eigenvector centrality of one.

Community 4 is famous for the spar. Among the official attractions, p142 (spar) and p317 (market) are included in this community. In addition, spar and hotel are included in the POI categories extracted based on weighted degree and eigenvector centrality. p142 has the second-highest weighted degree, and p317 is ranked 10th. As a different feature, community 5 has a higher proportion of bookstores among POI categories than other communities. p330 (bookstore) has the second-highest place based on weighted degree. In addition, the only arboretum belongs to the upper POI category. Moreover, as in community 1, there is a street where many visitors gather.

In community 6, several POIs under the cultural heritage category are included. However, compared to community 3, the overall value is low. p150 (cultural heritage), an official attraction belonging to this community, has the 33rd lowest weighted degree. Cafés and parks dominate the category in community 7. Based on the eigenvector centrality, p543(park) has the most significant value at 0.83758. In addition, p535(park), in the official attractions, also belongs to this community. Moreover, p535 is placed 23rd of 26 by weighted degree and has a relatively small value. In community 8, the park and café are distributed around p215 (forest), which has the highest weighted degree. Finally, in community 9, there are 16 categories, and POIs of various characteristics are distributed. Similar to community 7, extracted categories based on weighted degree and eigenvector centrality have a library.

Next, POIs excluded from the communities or the network are analyzed. We excluded 246 POIs, of which 156 POIs were connected to the network. Yuseong-Gu had 82 of 272 (30%), Daedeok-Gu 59 of 161 (37%), Dong-Gu 41 of 75 (56%), Jung-Gu 41 of 181 (23%), and Seo-Gu was 25 of 132 (19%) dropped POIs. In Dong-Gu, over half of the POIs were excluded from the network analysis. In addition, 22 of the 40 official attractions were overlooked, and the eigenvector centrality was low at 0.5 or less except for 6 POIs.

Fig. 11 shows the overall categories of POIs excluded in the network or community. Park accounts for the most significant proportion, followed by cultural heritages and forests. For parks and forests, a large site is required due to the characteristics of the categories. Moreover, there may be little connection as they are far away from other POIs. However, the significant number of POIs were excluded from the cultural heritage network despite being close to other POIs, or in the city center. p8, p21, and p312 were omitted from the network despite being official sites.

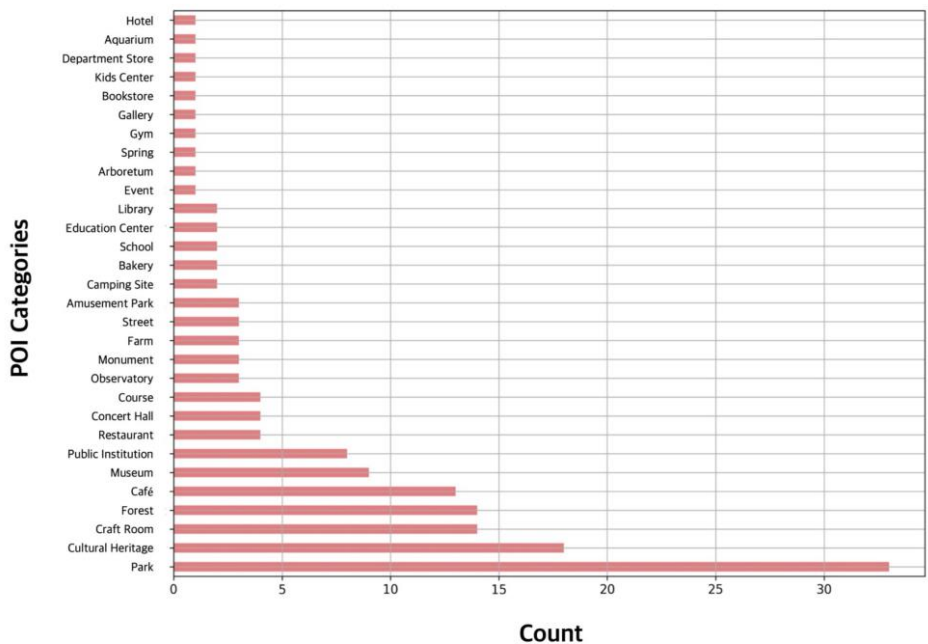

**Fig. 11** Distribution of the top 20 POI categories that do not belong to the communities and the network

## 5. Discussion

### 5.1. Relevance of analyzing POI

We investigated POI, including official attractions, from a community perspective. First, we analyzed influential POIs in the community and the official attractions list. Here, the influential POI signifies a relatively high weighted degree and a specific category within the community. The extracted POIs are places with many visitors. Moreover, the visitors can easily get much information, positively influencing the destination image (Beerli & Martín, 2004). Furthermore, since walkability is relatively high, visitors can experience a positive sense of place (Lo, 2009) while walking. Among official attractions, p142 (spar), p317 (local market), p146 (cultural heritage), p14 (station), and p138 (bakery) correspond to influential POIs (Table 11). Second, the POI that is not included in the community extraction can be considered far away in terms of walkability. Where the POI category requires a large site, such as a forest, it may be challenging to locate because of physical separation. However, given the number of visitors to the POI, there are places whose destination image needs to be increased. Notably, cultural heritage is not popular, except in community 3. Even in community 3, though the POIs of cultural heritage are physically well connected, influence is small, and places are unknown. Although cultural heritage accounts for 13% of the total POI, we conclude that cultural heritages in Daejeon are poorly recognized. These results lead to factors that help analyze visitors' perceptions of Daejeon.

**Table 11.** Classification of POI according to administrative districts

|  | POI Categories | POI | | | | |
|---|---|---|---|---|---|---|
|  |  | Yuseong-Gu | Daedeok-Gu | Dong-Gu | Jung-Gu | Seo-Gu |
| Influential POIs | Spar | p142 | - | - | - | - |
|  | Local Market | p317 | - | - | - | - |
|  | Bakery | p308 | - | - | p138 | - |
|  | Observatory | p169 | - | - | - | - |
|  | Cultural Heritage | p348 | p146 | - | - | - |
|  | Forest | p181 | - | - | - | - |
|  | Flower Shop | - | - | - | - | p761 |
|  | Street | - | - | - | p431 | - |
|  | Department Store | p292 | - | - | - | p692 |
|  | Bookstore | - | - | - | - | - |
|  | Library | p340 | p625, p596 | - | - | - |
|  | Concert Hall | - | - | - | - | p98 |
|  | Station | - | - | p14 | - | - |
|  | Craft room | - | p635 | - | - | - |
|  | Sum | 8 | 4 | 1 | 2 | 3 |
| Isolated POIs | School | p125 | - | - | - | - |
|  | Park | - | - | - | p135 | - |
|  | Observatory | p131 | - | - | - | - |
|  | Museum | - | - | - | p136 | - |
|  | Forest | - | - | - | - | p56 |
|  | Cultural Heritage | p312, p328 | p611, p614 | p8, p21 | p377, p386 | p94 |
|  | Sum | 4 | 2 | 2 | 4 | 2 |

Considering POI classification based on administrative districts, Yuseong-Gu has a large portion of POIs with strong influence, while Dong-Gu has low-impact POIs (Table 11). Currently, Daejeon city shares various courses for walking tours in Dong-Gu, but some parts are unrealistic. Although retrofitting may be necessary, some areas are difficult to access or unknown. The potential problems can only be solved by predicting and analyzing these problems in advance. In addition, although Daejeon City has 40 designated official attractions, few places are being visited. If the analysis results based on mobile-based visitor data can be reflected in the 40 official attractions, the possibility of devising a correct direction for a visit to Daejeon will increase.

*5.2. Visitor's walkability analysis method in the Korean context*

Daejeon, along with the research complex, is the fastest growing city in Korea. High-speed trains and bus terminals were activated, and vehicle-oriented roads were designed within the city, aiding its rapid growth. Investments to change Daejeon's transportation system for a people-centered community have been increasing. Though challenging, it is necessary to analyze the city structure with a data-driven method from the viewpoint of the visitor's walkability to understand the hidden context. We calculated the actual walking distance using 822 POIs and mobile-based visitor data in Daejeon for November 2021 and analyzed the results from a bottom-up perspective. The mobile-based visitor data is big data because it reflects the usage of the entire Daejeon city based on the accumulated visitors. For optimal results analysis, we converted the entire UTM-K-based data to WGS84 and calculated the distance between the POI and cells. Though a considerable amount of time is required for data preprocessing, it is relatively economical and time-efficient compared to the census conducted by government agencies. For census, each surveyor visits the house in person and takes more than an hour to survey each house. Because analysis is possible only after a certain amount of data has been accumulated, data must be disclosed quinquennially. In addition, accurately analyzing a specific context and place is challenging because the investigation is carried out in units of administrative districts. People visit cities based on information from specific places and not administrative districts. Therefore, we divided the cities based on the walkability through community detection to accurately analyze the visitor's situation and understand the city structure outside the administrative 'Gu.' Thus, a definition of a new division is necessary from the visitors' viewpoint, and an analysis method that understands the context of a specific city should be applied.

**6. Conclusion**

In this paper, we proposed an empirical network analysis method for the structural characteristics analysis of the city based on visitor's walkability in Daejeon, Korea. A structural understanding of the city is derived by quantitatively defining the relationship between the POI and cell. In addition, the analyzed POI-cell network contributes to understanding the urban context by focusing on specific places rather than census by viewing walkability with mobile-based visitor data. Furthermore, POI analysis results can be a key to enhancing the visitor experience, inducing revisit, and improving the destination image in newly extracted communities. However, this study has limitations as it uses centroid rather than boundary line of POI. In addition, mobile-based visitors are potential visitors who may go to a specific place, and only some of them may be actual visitors. In future work, the cell location will be specified using the extracted POI boundary data, and methods to improve proximity will be studied for an optimized experience. Proximity to a place can be an effective destination image enhancement and can be achieved by digitalization such as metaverse (Moreno et al., 2021). Therefore, the extracted POI will be implemented in a metaverse environment to study improvements in destination image and proximity of physically isolated places to overcome physical limitations with virtual proximity (Chang & Lee, 2021).